\documentclass[10pt]{iopart}

\usepackage{graphicx} 
\usepackage{color} 

\begin{document}

\title[Experimental investigation of kinetic instabilities driven by runaway electrons in the EXL-50 spherical torus]{Experimental investigation of kinetic instabilities driven by runaway electrons in the EXL-50 spherical torus}

\author{Mingyuan Wang$^{1,2,3}$, Mingsheng Tan$^{4}$\footnote{Authors to whom any correspondence should be addressed.}, Yuejiang Shi$^{2, 3}$\footnote{Authors to whom any correspondence should be addressed.}, Ziqi Wang$^{2, 3}$, Jiaqi Dong$^{2, 3}$, Adi Liu$^{5}$, Ge Zhuang$^{5}$, Songjian Li$^{2, 3}$, Shaodong Song$^{2, 3}$, Baoshan Yuan$^{2, 3}$, and Y-K Martin Peng$^{2, 3}$}

\address{$^1$ School of Mathematics and Physics, Anqing Normal University, Anqing, 246133, People's Republic of China}
\address{$^2$ Hebei Key Laboratory of Compact Fusion, Langfang 065001, People's Republic of China}
\address{$^3$ ENN Science and Technology Development Co., Ltd, Langfang 065001, People's Republic of China}
\address{$^4$ Institute of Energy, Hefei Comprehensive National Science Center, Hefei 230031, People's Republic of China}
\address{$^5$ Department of Plasma Physics and Fusion Engineering, University of Science and Technology of China, Hefei 230026, People's Republic of China}

\ead{tanms@ie.ah.cn and yjshi@ipp.ac.cn}
\vspace{10pt}

\begin{indented}
\item[]July 2023
\end{indented}

\begin{abstract}
In this study, the first observation of high-frequency instabilities driven by runaway electrons has been reported in the EXL-50 spherical torus using a high-frequency magnetic pickup coil. The central frequency of these instabilities is found to be exponentially dependent on the plasma density, similar to the dispersion relation of the whistler wave. The instability frequency displays chirping characteristics consistent with the Berk-Breizman model of beam instability. Theoretically, the excitation threshold of the instability driven by runaway electrons is related to the ratio of the runaway electron density to the background plasma density, and such a relationship is first demonstrated experimentally in this study. The instability can be stabilized by increasing the plasma density, consistent with the wave-particle resonance mechanism. This investigation demonstrates the controlled excitation of chirping instabilities in a tokamak plasma and reveals new features of these instabilities, thereby advancing the understanding of the mechanisms for controlling and mitigating runaway electrons.
\end{abstract}
 
\vspace{2pc}
\noindent{\it Keywords}: runaway electrons, frequency chirping, kinetic instabilities, runaway electron mitigation.

\submitto{\NF}
\maketitle
\ioptwocol

\section{Introduction}

In plasmas related to nuclear fusion, energetic particles that are significantly hotter than the plasma bulk can be generated through various mechanisms, such as auxiliary heating (electron cyclotron resonance heating (ECRH), ion cyclotron resonance heating (ICRH) and neutral beam injection (NBI)), alpha particles\cite{fasoli2007chapter} or runaway avalanche\cite{rosenbluth1997theory}. During plasma disruptions, the current quench (CQ) induces a large toroidal electric field, which accelerates runaway electrons (REs) and produces a relativistic beam. In the process of slowing down of energetic particles, kinetic instabilities driven by them can be generated through nonlinear wave-particle interactions (WPIs).

One of the primary threats to the stable operation of tokamak fusion reactors is the transfer of the plasma current from thermal to relativistic electrons during CQ\cite{lehnen2015disruptions}. REs can cause serious localized damage to the plasma-facing components (PFCs)\cite{boozer2018pivotal}. Numerous theoretical and experimental research efforts have been dedicated to developing effective mitigation methods for these effects, particularly for future fusion reactors. A simple and fundamental mitigation strategy is to inject impurities into the post-disruption plasma to increase radiation, which is expected to be used in ITER. Other mitigation strategies under investigation include magnetic perturbations generated by external magnetic coils\cite{chen2018suppression}, as well as WPIs driven by either kinetic instabilities\cite{liu2018role} or external wave injection\cite{guo2018control}.

The interaction of plasma waves or instabilities with energetic particles can cause the redistribution and loss of energetic particles. The resulting instabilities enhance the transport of energetic particles, and energetic particles escape from the confinement region, which causes additional particle energy loss. Therefore, appropriate instability may favor the mitigation of REs, as observed in the DIII-D tokamak\cite{lvovskiy2019observation}.

Kinetic instabilities driven by energetic particles have two forms, one involves the evolution of mode frequency with the plasma equilibrium parameters, and the other is the frequency chirping. Because the instabilities that redistribute energetic particles are generated by the particles themselves, the process is inherently nonlinear. Relativistic REs, being high-energy beams, are capable of producing chirping instabilities.

In experiments, Alfv\'en eigenmodes and whistler waves excited by energetic particles are observed. The frequency of steady-frequency instabilities is quasi-monochromatic, and its parametric behavior in a cold plasma can be expressed as \cite{fulop2006destabilization, spong2018first, heidbrink2019low},
\begin{equation}\label{equ:whis_wave_freq}
  \omega = k v_A \sqrt{1 + k_\parallel^2 c^2 / \omega_{pi}^2} ,
\end{equation}
where $k$ represents the wavenumber, $v_A$ denotes the Alfv\'en velocity, $k_\parallel$ is the wavenumber parallel to the magnetic field, $c$ is the speed of light, and $\omega_{pi}$ represents the ion plasma frequency, respectively. The Alfv\'en velocity is given by $v_A = B / \sqrt{\mu_0 \rho}$, where $B$ is the magnetic field strength, $\mu_0$ is the permeability of vacuum, and $\rho$ is the plasma mass density ($\rho \propto n_e$). Therefore, it scales inversely with the square root of plasma density ($v_A \propto n_e^{-1/2}$). From Eq. (\ref{equ:whis_wave_freq}), the instability frequency scales with the plasma density, ranging from $n_e^{-1/2}$ ($k_\parallel^2 c^2 \ll \omega_{pi}^2$) to $n_e^{-1}$ ($k_\parallel^2 c^2 \gg \omega_{pi}^2$)\cite{spong2018first}.

A commonly used analytical theory for beam instability is the Berk-Breizman model, which demonstrates various nonlinear behaviors in steady-state dynamics. In a dissipative plasma with a large number of energetic particles, nonlinear WPIs produce self-trapped structures (holes and clumps) in the phase-space. The presence of a particle source can also affect the nonlinear behavior. Typically, chirping instabilities exhibit two branches of frequency, one increasing upward and the other decreasing downward. In the Berk-Breizman model, chirping instabilities spontaneously burst with a frequency shift and time relationship of $\delta \omega \propto \sqrt{t}$\cite{berk1997spontaneous, berk1999spontaneous}. The effect of various collision operators on the nonlinear frequency chirping of fast-ion-driven toroidicity-induced Alfv\'en eigenmodes (TAEs) has been studied in tokamak and Wendelstein 7-X (W7-X)\cite{slaby2019on}. The Krook operator is associated with periodic, well-separated chirping events that are routinely observed in experiments and numerical simulations. If the diffusion effect increases, the motion of hole-clump pairs can be suppressed, leading to the gradual disappearance of sub-branches\cite{hou2021nonlinear}. Additionally, drag can generate asymmetric frequency chirping, which is necessary to enhance holes and suppress clumps.

Relaxation oscillations accompanied by quasi-periodic chirping instabilities burst have been observed in various devices with typical periods in the order of milliseconds\cite{valovic2000quasi, fredrickson2006collective, gryaznevich2006perturbative, gryaznevich2008recent, bierwage2017self, melnikov2018detection, lvovskiy2019observation}. In fusion experiments, chirping instabilities are widespread in various frequency bands and exhibit different frequency structures, such as chirping-up frequency, chirping-down frequency and chirping-bifurcating frequency.

These instabilities are excited when REs resonate with the wave. The wave phase seen by REs remains stationary when\cite{spong2018first, heidbrink2019low},
\begin{equation}\label{equ:ome_seen_wave}
  \omega - k_\parallel v_\parallel - k_\perp v_d - l \Omega_{ce} / \gamma = 0 ,
\end{equation}
where $\omega$ is the wave frequency, $v_\parallel$ is the parallel velocity, $k_\perp$ is the perpendicular component of the wavenumber, $v_d$ is the orbital drift, $\Omega_{ce}$ is the electron cyclotron frequency, and $\gamma$ is the relativistic factor, respectively. Here, $l$ is an integer, and the main excitation components are $l = 0 , \pm 1$. If $\Omega_{ce} >  0$, then the cases of $l = -1$, $l = 0$ and $l = +1$ correspond to the anomalous Doppler, Cherenkov and normal Doppler resonances, respectively. Theoretically, resonance may occur at any value of the integer $l$, but in tokamak experiments, the measured wave frequency is much smaller than the electron cyclotron frequency ($\omega \ll \Omega_{ce}$), so these resonant conditions are fulfilled when $l = -1$ or $l = 0$.

The threshold for magnetosonic-whistler instability driven by relativistic REs is expressed as\cite{fulop2006destabilization, fulop2009magnetic},
\begin{equation}\label{equ:inst_thre}
  \frac{n_r}{n_e} > \frac{Z^2 B_T}{20 T_{eV}^{3/2}} ,
\end{equation}
where $Z$ represents the effective ion charge number, $B_T$ is the magnetic field strength in unit of T, $T_{eV}$ is the background plasma temperature in unit of eV, $n_r$ and $n_e$ refer to the RE density and background plasma density, respectively. Eq. (\ref{equ:inst_thre}) implies that collisional damping has less impact on high temperature plasmas, resulting in a lower density of REs required to drive the instability. This threshold is independent of plasma density and is only relevant for the fraction of REs at low magnetic field strengths. The temporal evolution of this threshold is highly sensitive to various plasma parameters.

This study presents the first observation of kinetic instabilities driven by REs in the XuanLong-50 (EXL-50) spherical torus\cite{shi2022solenoid}. This paper is organized as follows. In Section \ref{sec_exp_diag}, the EXL-50 experiment and the diagnostics used in this study are introduced. The characteristics of the observed kinetic instabilities are described in Section \ref{obse_freq_chir_inst}, including dispersion relation, chirping characteristic, resonant condition and excitation threshold. Finally, the conclusion is given in Section \ref{disc_conc}.

\section{Experiment and diagnostics}
\label{sec_exp_diag}

The experiments are carried out on the EXL-50 spherical tokamak at the Energy iNNovation (ENN) science and technology development private limited company in Langfang, China. Hydrogen plasmas are used and ECRH is employed to heat plasma and drive the plasma current. The EXL-50 is a medium-sized spherical tokamak with a major radius of $R_0 = 0.58$ m and a minor radius of $a = 0.41$ m, and it has been in operation since early 2020\cite{shi2022solenoid}. The toroidal field $B_t$ can reach a maximum of approximately 0.5 T at $r \approx 0.48$ m, and the aspect ratio $A$ is smaller than 2.

Three sets of ECRH systems (28 GHz) are used and can produce a large number of energetic electrons in the EXL-50 spherical tokamak. Discharges of the plasma current substantially above 100 kA can be obtained by 100 kW ECRH. In this study, only ECRH is used to heat and maintain the plasma, with the plasma current kept at $I_p$ = 50-100 kA and the line-integrated plasma density at $\int {n_e dl}$ = (1-4)$\times 10^{18}$ m$^{-2}$.

\begin{figure}[htbp]
  \centering
  \includegraphics[width=0.35\paperwidth]{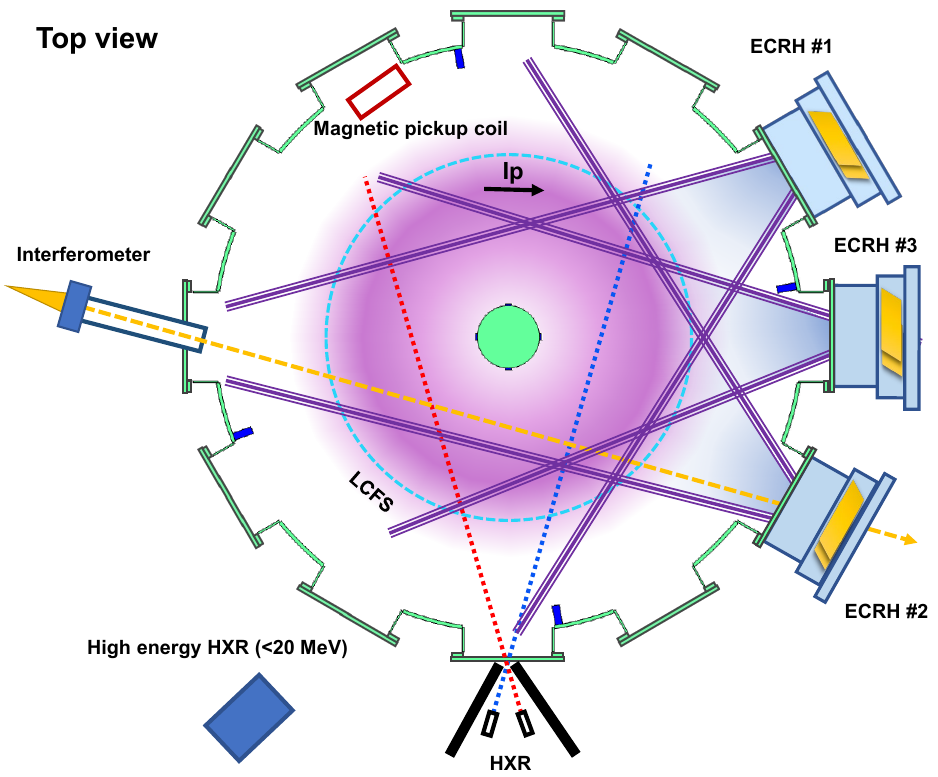}\\
  \caption{A top view of the diagnostic layout of the EXL-50 spherical tokamak.}
  \label{fig:diag_layout}
\end{figure}

The diagnostic layout of the EXL-50 spherical tokamak used in this study is illustrated in Fig. \ref{fig:diag_layout}. A microwave interferometer operated at 140 GHz is used to measure the line-integrated plasma density\cite{li2021quasi}. A distant hard X-ray (HXR) scintillating plastic detector is placed a few meters away from the EXL-50 machine to detect the volumetric bremsstrahlung emission resulting from the loss of energetic electrons to the machine wall. The interaction between energetic electrons lost to the wall and the wall material causes a rapid increase of this measured signal. A high-frequency magnetic pickup coil mounted at the low-field side (LFS) on the mid-plane of the vacuum vessel is used to diagnose the frequency chirping instabilities\cite{wang2023radio}. The coil has a frequency response from 1 MHz to 200 MHz and is digitized by a digital oscilloscope with a maximum sampling rate of 3 GS/s.

\section{Characteristics of kinetic instabilities driven by runaway electrons}
\label{obse_freq_chir_inst}

\begin{figure}[htbp]
  \centering
  \includegraphics[width=0.35\paperwidth]{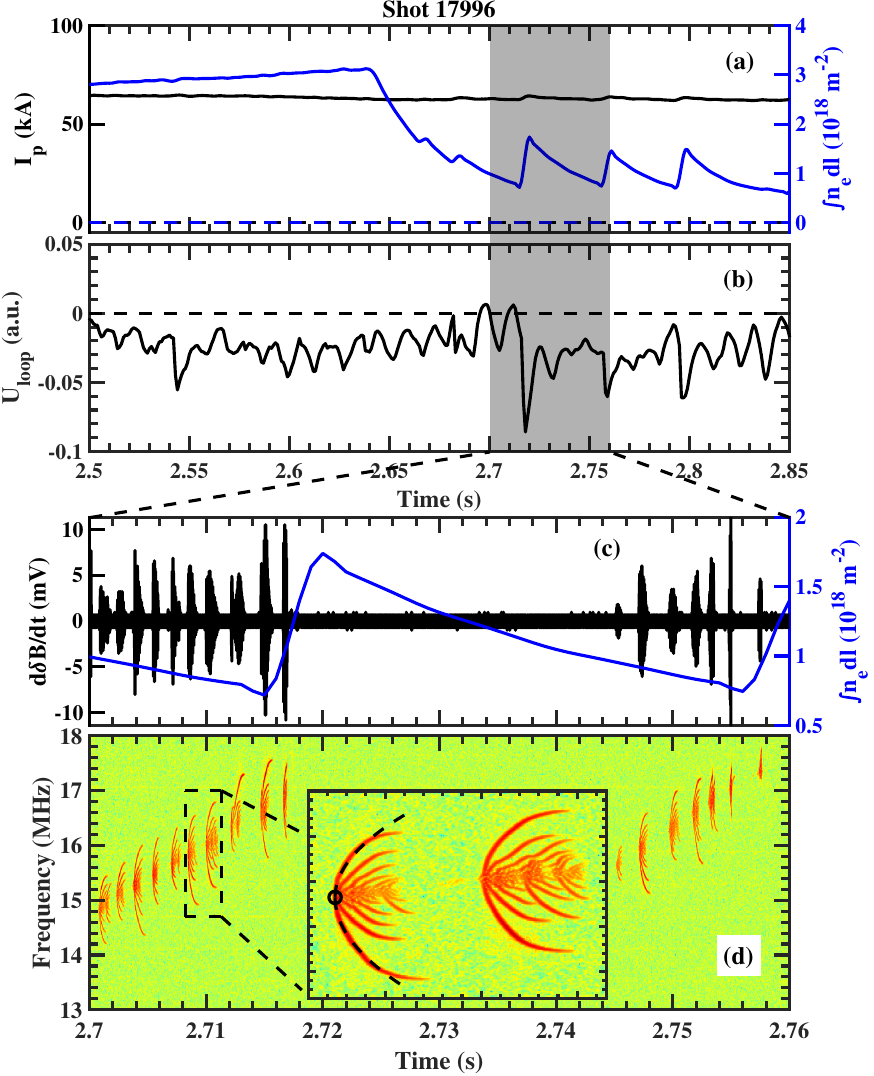}\\
  \caption{Time evolution of plasma parameters: (a) plasma current, plasma integral density, (b) loop voltage and HXR signal. Signals in the time of the shadow: (c) raw signal measured by the high-frequency magnetic pickup coil, plasma integral density, and (d) spectrogram of the raw signal.}
  \label{fig:basic_para_bdata_chirp_freq}
\end{figure}

Fig. \ref{fig:basic_para_bdata_chirp_freq} shows the plasma parameters and raw signal measured by the magnetic pickup coil obtained from a typical ECRH discharge. The plasma is ignited and sustained using two sets of ECRH systems. The loop voltage is close to zero during the flat-top of the plasma current. For the normal discharges in EXL-50, the plasma current is carried by the energetic electrons produced by ECRH\cite{shi2022solenoid}. The velocity distribution of energetic electrons driven by ECRH is different to that of runaway electrons induced by a toroidal electric field. In the former case, the energetic electrons possess similar magnitudes of parallel and perpendicular velocities, while in the latter, the parallel velocities dominate. On the other hand, considerable toroidal electric field can be induced by internal reconnection event (IRE) or disruption in EXL-50. Part of energetic electrons is converted to runaway electrons during IRE or disruption phase which will excite special instabilities. When the second ECRH system is shut down, the plasma density drops sharply, and there are several relaxation oscillations in the line-integrated plasma density during 2.7-3 s. There are many high frequency fluctuations in the raw signal data from the magnetic pickup coil.

The frequency of chirping instabilities can be obtained from the spectrogram of the magnetic field measured by the high-frequency magnetic pickup coil. The time evolution of a detailed frequency of kinetic instabilities is shown in Fig. \ref{fig:basic_para_bdata_chirp_freq}(d) during the time of 2.7-2.76 s as shown in the shadow of Fig. \ref{fig:basic_para_bdata_chirp_freq}(a), and it consists of two components, $\omega \left( t \right) = \omega_0 \left( t \right) + \delta \omega \left( t \right)$, where $\omega_0 \left( t \right)$ and $\delta \omega \left( t \right)$ are the central frequency and the time-varying frequency part of each burst cycle, respectively. Obviously, the central frequency of chirping instabilities depends on the line-integrated plasma density, which will be described in detail later. The frequency of kinetic instabilities is driven by REs, and during this period, the instability frequency varies within a range of 14-18 MHz.

\subsection{Dispersion relation}

\begin{figure*}[htbp]
  \centering
  \includegraphics[width=0.7\paperwidth]{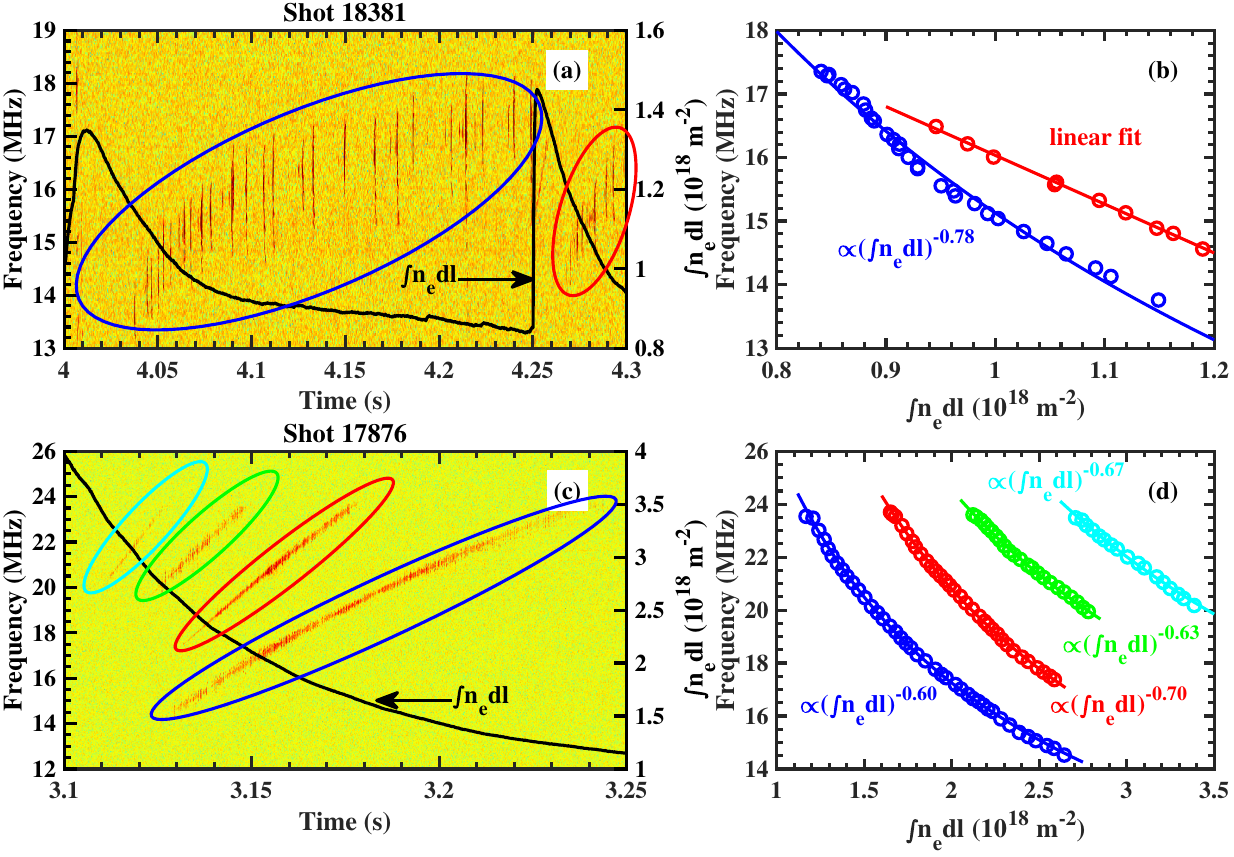}\\
  \caption{The dispersion relation of chirping instabilities: (a), (c) the time evolution of chirping frequencies and the line-integrated plasma density, (b) and (d) the relation between the central frequency and the line-integrated plasma density.}
  \label{fig:plas_freq_vs_nel_1}
\end{figure*}

The central frequency of chirping instabilities increases as the plasma density decreases, indicating an inverse correlation between them. The dependence of the central frequency of chirping instabilities on the line-integrated plasma density, $f \left( \int{n_e dl} \right)$, is shown in Fig. \ref{fig:plas_freq_vs_nel_1}. In Fig. \ref{fig:plas_freq_vs_nel_1}(a), the line-integrated plasma density, $\int{n_e dl}$, exhibits a quasi-periodic oscillation, increasing rapidly and then decreases slowly.

The parametric behavior of the central frequency of chirping instabilities can be understood through the cold plasma dispersion relation given by Eq. (\ref{equ:whis_wave_freq}). In the lowest order case, $\omega = k v_A$, this predicts that a linear scaling of frequency $\omega$ with magnetic field strength and an inverse square root scaling with plasma density. The dispersion relation of chirping instabilities is illustrated in Fig. \ref{fig:plas_freq_vs_nel_1}. The central frequency of instabilities has a relation with the line-integrated plasma density in the form of ${\left( \int{n_e dl} \right)^{-b}}$, with the exponent $b$ ranging from 0.5 to 1. Therefore, from the experimental results, the exponent $b$ is consistent with the theoretical results. Chirping instabilities are also observed at higher frequencies in the range of several tens of MHz.

\begin{figure*}[htbp]
  \centering
  \includegraphics[width=0.7\paperwidth]{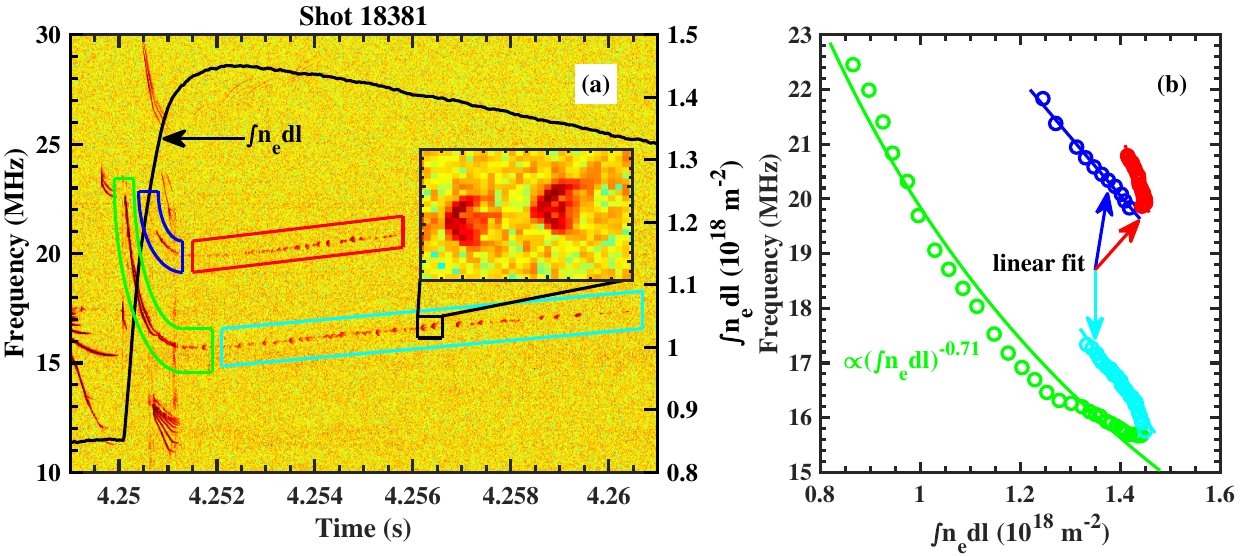}\\
  \caption{The dispersion relation of chirping instabilities: (a) the time evolution of chirping frequencies and the line-integrated plasma density, and (b) the relation between the central frequency and the line-integrated plasma density.}
  \label{fig:plas_freq_vs_nel_2}
\end{figure*}

The small-scale frequency chirping depicted in the inset of Fig. \ref{fig:plas_freq_vs_nel_2}(a) highlights another example of frequency modulation at relatively small time and frequency scales. Fig. \ref{fig:plas_freq_vs_nel_2} illustrates multiple frequency bands associated with chirping instabilities whose central frequencies are inversely proportional to the line-integrated plasma density. As the plasma density rapidly increases, the lower branches of the large-scale frequency chirping instabilities start appearing, indicated by the blue and green boxes. Conversely, as the plasma density slowly decreases, the large-scale frequency chirping instabilities reappear, depicted by the red and cyan boxes. Interestingly, these branches display small-scale frequency chirping instabilities, as revealed in the inset of  Fig. \ref{fig:plas_freq_vs_nel_2}(a). The observed relation between the instability frequency and the plasma density satisfies $f \propto \left( \int{n_e dl} \right)^{-b}$, which is consistent with the dispersion relation of whistler wave.

It is worth noting that when the plasma density increases rapidly, the central frequency changes rapidly with the plasma desity, so it is coupled with the rapidly changing chirping frequency. This causes the two frequencies to be difficult to distinguish, so that the relation between the central frequency and plasma density is not well fitted, as shown in the blue and green parts of Fig. \ref{fig:plas_freq_vs_nel_2}. In the red and cyan parts, the plasma density changes slowly, so the central frequency changes less, and it is easier to separate from the rapidly changing chirping frequency.

\subsection{Chirping characteristic}

The instability bursts are quasi-periodic with a burst period of a few milliseconds, such as 1.56 ms as shown in the inset of Fig. \ref{fig:basic_para_bdata_chirp_freq}(d). The chirping frequency $\delta \omega \left( t \right)$ of each burst cycle has two branches - an upward-chirping frequency and a downward-chirping frequency - they are slightly asymmetric. The instability frequency of a burst cycle in the inset of Fig. \ref{fig:basic_para_bdata_chirp_freq}(d) varies over a frequency range of 14.8 MHz to 16.6 MHz over a duration of 1 ms.

The black dashed line in the inset of Fig. \ref{fig:basic_para_bdata_chirp_freq}(d) shows the relation between the chirping frequency and the square root of time, $\delta \omega \left( t \right) \propto \pm \sqrt{t}$, and the black circle shows the central frequency $\omega_0 \left( t \right)$. The relationship between the chirping frequency and time is consistent with the theoretical result predicted by the Berk-Breizman paradigm, and there is a plasma density threshold for the excitation of instabilities.

Two bursts of chirping instabilities are magnified in the inset of Fig. \ref{fig:basic_para_bdata_chirp_freq}(d), where the first burst chirps both downward and upward in the frequency range of about 1.8 MHz, or $\delta \omega / \omega_0 \approx  11.3 \%$. There is also a series of secondary instability frequencies inside the main chirping instabilities, which may be excited by changes in the electron energy distribution produced by the main instability. The downward chirping component lasts longer, but has a smaller amplitude than that of the upward chirping component. 

\begin{figure*}[htbp]
  \centering
  \includegraphics[width=0.7\paperwidth]{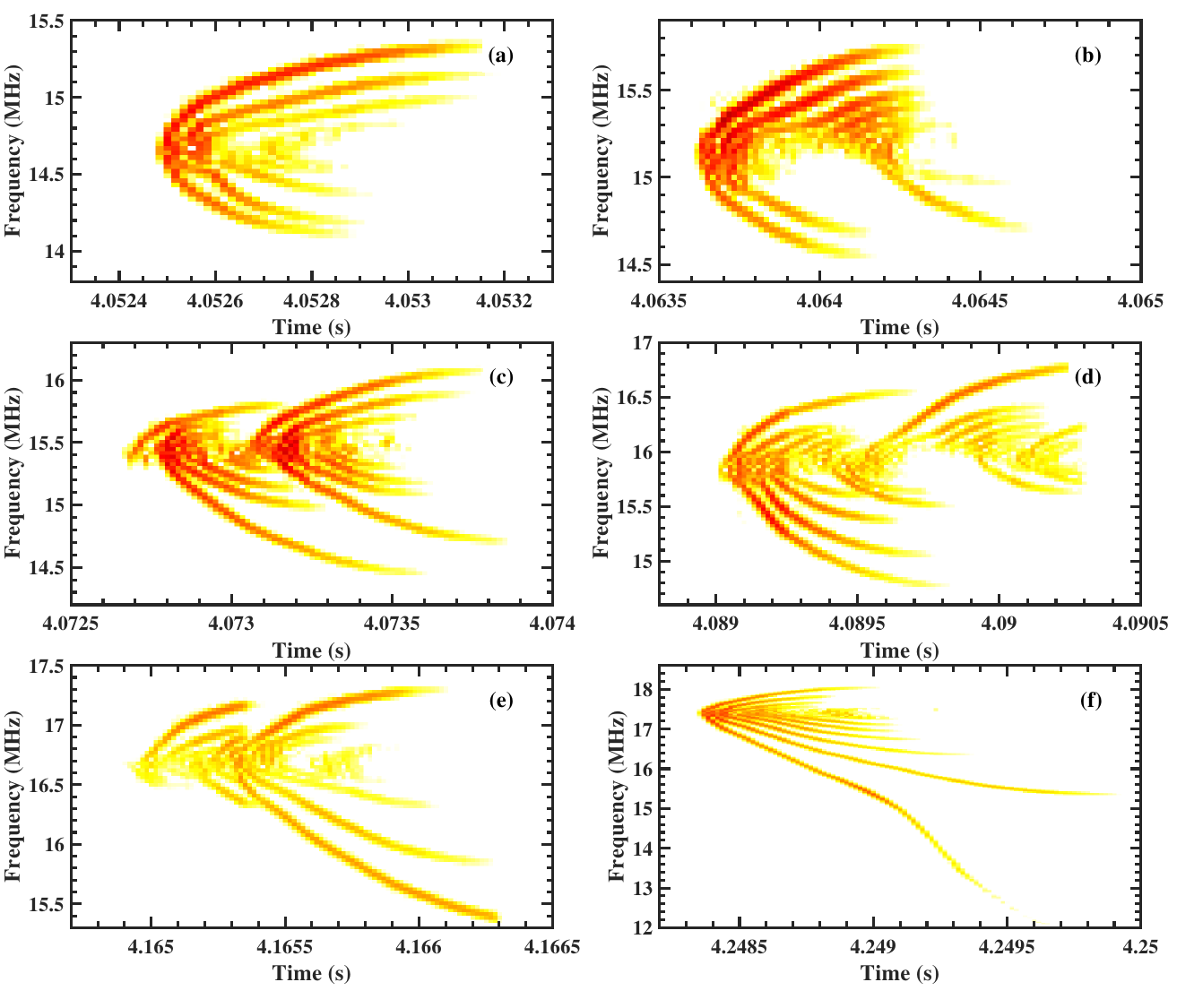}\\
  \caption{The different chirping frequency structures of Shot 18381.}
  \label{fig:shot18381_sp_chir_mode}
\end{figure*}

The chirping frequencies have various structures, as shown in Fig. \ref{fig:shot18381_sp_chir_mode}, which may have different collision operators and be related to different theoretical mechanisms. The drag effect can form asymmetric chirping frequencies, which enhance holes and suppress clumps, and in experiments, most of the observed chirping frequencies are asymmetric as shown in Fig. \ref{fig:shot18381_sp_chir_mode}.

\subsection{Resonance condition}

\begin{figure}[htbp]
  \centering
  \includegraphics[width=0.35\paperwidth]{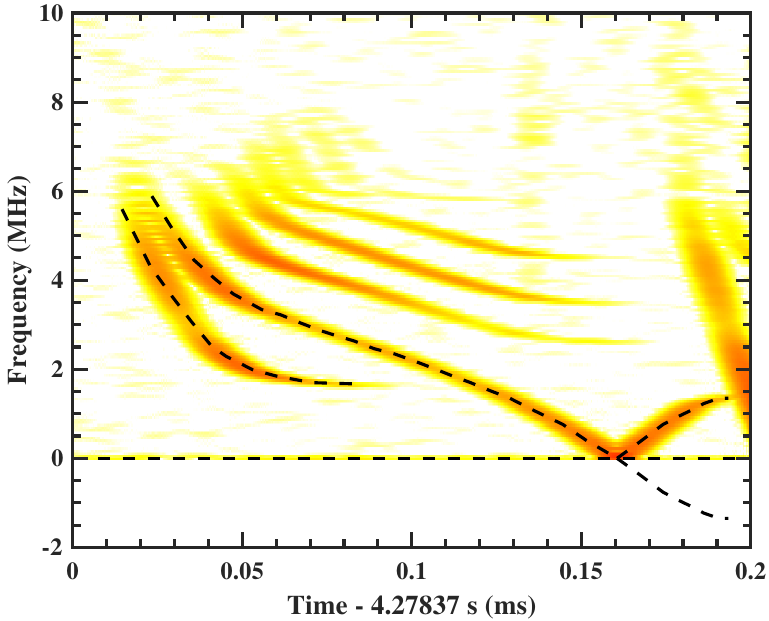}\\
  \caption{The chirping frequency evolving in time crossing zero-frequency.}
  \label{fig:over_zero_mode}
\end{figure}

Chirping instabilities arise from the resonate of energetic electrons with wave. As shown in Eq. (\ref{equ:ome_seen_wave}), these instabilities can exhibit negative frequencies when $l = -1$ due to anomalous Doppler resonance. In the case of the EXL-50 experiment, the observed wave frequency is much lower than the electron resonance frequency, allowing for the observation of resonances occurring at $l = -1$ or $l = 0$ in the low frequency band. When $l = -1$, the chirping frequency exhibits time-dependent evolution and may cross zero-frequency, as depicted in Fig. \ref{fig:over_zero_mode}. This frequency evolution consists of multiple lower branches, one of which lasts longer and continues to evolve across zero-frequency.

\subsection{Excitation threshold of instability}

\begin{figure}[htbp]
  \centering
  \includegraphics[width=0.35\paperwidth]{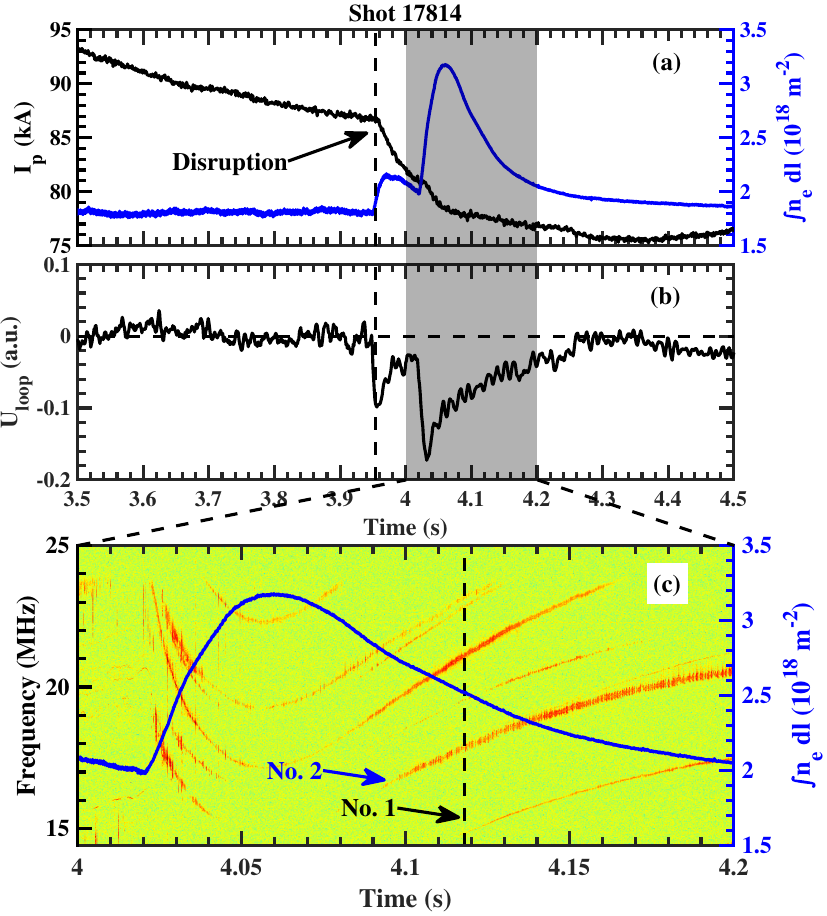}\\
  \caption{Discharge parameters during a disruption: (a) the plasma current, the line-integrated plasma density, and (b) the loop voltage. (c) The time evolution of chirping frequency when instabilities are stabilized.}
  \label{fig:Ip_nel_Uloop_freq_thre}
\end{figure}

Fig. \ref{fig:Ip_nel_Uloop_freq_thre} displays the time evolution of the plasma parameters during a disruption event. As the plasma current decreases rapidly, a corresponding loop voltage is induced and the plasma density increases rapidly. This, in turn, causes the acceleration of electrons by the resulting toroidal electric field, ultimately resulting in the formation of relativistic REs.

\begin{figure}[htbp]
  \centering
  \includegraphics[width=0.35\paperwidth]{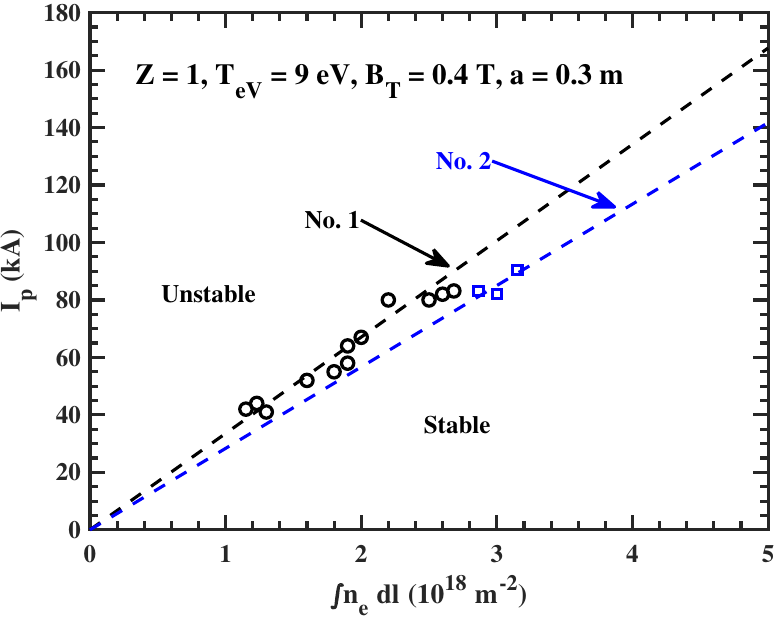}\\
  \caption{The points are the critical values of the plasma current and the line-integrated plasma density when instabilities are stabilized. The black dashed line indicates the relation of Eq. (\ref{equ:Ip_nedl_thres}) with parameters $Z$ = 3, $T_{eV}$ = 20 eV, $B_T$ = 0.4 T and $a$ = 0.11 m.}
  \label{fig:nel_Ip_linear_fitting}
\end{figure}

It is evident from Fig. \ref{fig:Ip_nel_Uloop_freq_thre}(c) that when the plasma density surpasses a certain threshold, chirping instabilities no longer appear. Additionally, chirping instabilities have multiple harmonic components, each with a distinct excitation threshold. Assuming that the minor radius of REs is $a$, $n_e \gg n_r$, and the current of relativistic REs is dominant, $j \approx e n_r c$, so $n_r / n_e \approx j / e n_e c$, then Eq. (\ref{equ:inst_thre}) can be rephrased as,
\begin{equation}\label{equ:Ip_nedl_thres}
  \frac{I_p}{\int{n_e dl}} > \frac{ \pi e c a Z^2 B_T}{20 T_{eV}^{3/2}} .
\end{equation}
Critical values of the plasma current and line-integrated plasma density required for stabilizing whistler instabilities are represented by points in Fig. \ref{fig:nel_Ip_linear_fitting}, where different colors and marks indicate different harmonics. For instance, black circles indicate the frequency band of No. 1. The black dashed line corresponds to the relation given by Eq. (\ref{equ:Ip_nedl_thres}) with parameters $Z$ = 3, $T_{eV}$ = 20 eV, $B_T$ = 0.4 T and $a$ = 0.11 m. The cases with higher-order harmonic components are represented by blue squares. The instability can be stabilized by increasing the plasma density, which is consistent with the wave-particle resonance mechanism.

\section{Conclusion}
\label{disc_conc}

For the first time on the EXL-50 spherical tokamak, frequency chirping instabilities driven by energetic electrons have been observed. These instabilities are identified using a high-frequency magnetic pickup coil mounted on the vacuum vessel. They are detected in a plasma driven by the ECRH and associated with energetic electrons during the ECRH steady-state plasma.

The central frequency of kinetic instabilities is exponentially related to the line-integrated plasma density, $f \propto \left( \int {n_e dl} \right)^{-b}$, where the exponent $b$ is between 0.5 to 1, in agreement with Alfv\'enic scaling. Although the frequency chirps by 14.8-16.6 MHz on a timescale of 1 ms, there are numerous frequency bands in the EXL-50 spherical tokamak. The measured instability frequency structures are consistent with the hole-clump model for frequency chirping driven by energetic electrons. From the resonance condition, the chirping instabilities have negative frequencies, which has been experimentally observed. Theoretically, the excitation threshold of the whistler instability driven by REs is related to the ratio of the RE density to the background plasma density, and such a relationship is first demonstrated experimentally in this study. The relationship between the plasma current and plasma density during instability bursts is in agreement with that predicted by the theoretical model of the whistler instability threshold excited by REs. The instability can be stabilized by increasing the plasma density, which is consistent with the wave-particle resonance mechanism. This research demonstrates the controlled excitation of chirping instabilities in a tokamak plasma and describes several new features, which include the dispersion relation, the chirping characteristic and the resonance condition. It improves the understanding of the physical picture of RE control and mitigation.

\ack
This work is supported by the High-End Talents Program of Hebei Province, Innovative Approaches towards Development of Carbon-Free Clean Fusion Energy (No. 2021HBQZYCSB006). The authors gratefully acknowledge the support of the Institute of Energy, Hefei Comprehensive National Science Center (No. 21KZS202).

\section*{References}
\bibliographystyle{iopart-num}
\bibliography{cite_paper}

\providecommand{\newblock}{}
\begin{thebibliography}{10}
\expandafter\ifx\csname url\endcsname\relax
  \def\url#1{{\tt #1}}\fi
\expandafter\ifx\csname urlprefix\endcsname\relax\def\urlprefix{URL }\fi
\providecommand{\eprint}[2][]{\url{#2}}

\bibitem{fasoli2007chapter}
Fasoli A, Gormenzano C, Berk H~L, Breizman B, Briguglio S, Darrow D~S,
  Gorelenkov N, Heidbrink W~W, Jaun A, Konovalov S~V, Nazikian R, Noterdaeme
  J~M, Sharapov S, Shinohara K, Testa D, Tobita K, Todo Y, Vlad G and Zonca F
  2007 {\em Nuclear Fusion\/} {\bf 47} S264--S284
  \urlprefix\url{https://dx.doi.org/10.1088/0029-5515/47/6/S05}

\bibitem{rosenbluth1997theory}
Rosenbluth M~N and Putvinski S~V 1997 {\em Nuclear Fusion\/} {\bf 37}
  1355--1362 \urlprefix\url{https://dx.doi.org/10.1088/0029-5515/37/10/I03}

\bibitem{lehnen2015disruptions}
Lehnen M, Aleynikova K, Aleynikov P~B, Campbell D~J, Drewelow P, Eidietis N~W,
  Gasparyan Y, Granetz R~S, Gribov Y, Hartmann N, Hollmann E~M, Izzo V~A,
  Jachmich S, Kim S~H, Ko\v{c}an M, Koslowski H~R, Kovalenko D, Kruezi U,
  Loarte A, Maruyama S, Matthews G~F, Parks P~B, Pautasso G, Pitts R~A, Reux C,
  Riccardo V, Roccella R, Snipes J~A, Thornton A~J, de~Vries P~C and {EFDA JET
  contributors} 2015 {\em Journal of Nuclear Materials\/} {\bf 463} 39--48
  \urlprefix\url{https://www.sciencedirect.com/science/article/pii/S0022311514007594}

\bibitem{boozer2018pivotal}
Boozer A~H 2018 {\em Nuclear Fusion\/} {\bf 58} 036006
  \urlprefix\url{https://dx.doi.org/10.1088/1741-4326/aaa1db}

\bibitem{chen2018suppression}
Chen Z~Y, Lin Z~F, Huang D~W, Tong R~H, Hu Q~M, Wei Y~N, Yan W, Dai A~J, Zhang
  X~Q, Rao B, Yang Z~J, Gao L, Dong Y~B, Zeng L, Ding Y~H, Wang Z~J, Zhang M,
  Zhuang G, Liang Y, Pan Y, Jiang Z~H and {J-TEXT Team} 2018 {\em Nuclear
  Fusion\/} {\bf 58} 082002
  \urlprefix\url{https://dx.doi.org/10.1088/1741-4326/aab2fc}

\bibitem{liu2018role}
Liu C, Hirvijoki E, Fu G~Y, Brennan D~P, Bhattacharjee A and Paz-Soldan C 2018
  {\em Physical Review Letters\/} {\bf 120} 265001
  \urlprefix\url{https://www.ncbi.nlm.nih.gov/pubmed/30004735}

\bibitem{guo2018control}
Guo Z, McDevitt C~J and Tang X~Z 2018 {\em Physics of Plasmas\/} {\bf 25}
  032504 \urlprefix\url{https://doi.org/10.1063/1.5019381}

\bibitem{lvovskiy2019observation}
Lvovskiy A, Heidbrink W~W, Paz-Soldan C, Spong D~A, Dal~Molin A, Eidietis N~W,
  Nocente M, Shiraki D and Thome K~E 2019 {\em Nuclear Fusion\/} {\bf 59}
  124004 \urlprefix\url{https://dx.doi.org/10.1088/1741-4326/ab4405}

\bibitem{fulop2006destabilization}
F\"{u}l\"{o}p T, Pokol G, Helander P and Lisak M 2006 {\em Physics of
  Plasmas\/} {\bf 13} 062506 \urlprefix\url{https://doi.org/10.1063/1.2208327}

\bibitem{spong2018first}
Spong D~A, Heidbrink W~W, Paz-Soldan C, Du X~D, Thome K~E, Van~Zeeland M~A,
  Collins C, Lvovskiy A, Moyer R~A, Austin M~E, Brennan D~P, Liu C, Jaeger E~F
  and Lau C 2018 {\em Physical Review Letters\/} {\bf 120} 155002
  \urlprefix\url{https://link.aps.org/doi/10.1103/PhysRevLett.120.155002}

\bibitem{heidbrink2019low}
Heidbrink W~W, Paz-Soldan C, Spong D~A, Du X~D, Thome K~E, Austin M~E, Lvovskiy
  A, Moyer R~A, Pinsker R~I and Van~Zeeland M~A 2019 {\em Plasma Physics and
  Controlled Fusion\/} {\bf 61} 014007
  \urlprefix\url{https://dx.doi.org/10.1088/1361-6587/aae2da}

\bibitem{berk1997spontaneous}
Berk H~L, Breizman B~N and Petviashvili N~V 1997 {\em Physics Letters A\/} {\bf
  234} 213--218
  \urlprefix\url{https://www.sciencedirect.com/science/article/pii/S0375960197005239}

\bibitem{berk1999spontaneous}
Berk H~L, Breizman B~N, Candy J, Pekker M and Petviashvili N~V 1999 {\em
  Physics of Plasmas\/} {\bf 6} 3102--3113
  \urlprefix\url{https://doi.org/10.1063/1.873550}

\bibitem{slaby2019on}
Slaby C, K\"{o}nies A, Kleiber R and Leyh H 2019 {\em Nuclear Fusion\/} {\bf
  59} 046006 \urlprefix\url{https://dx.doi.org/10.1088/1741-4326/aafe31}

\bibitem{hou2021nonlinear}
Hou Y, Chen W, Yu L, Zou Y, Xu M and Duan X 2021 {\em Chinese Physics
  Letters\/} {\bf 38} 045202
  \urlprefix\url{https://dx.doi.org/10.1088/0256-307X/38/4/045202}

\bibitem{valovic2000quasi}
Valovi\v{c} M, Lloyd B, McClements K~G, Warrick C~D, Fielding S~J, Morris A~W,
  Pinfold T, Wilson H~R, {COMPASS-D Team} and {ECRH Team} 2000 {\em Nuclear
  Fusion\/} {\bf 40} 1569--1573
  \urlprefix\url{https://dx.doi.org/10.1088/0029-5515/40/9/101}

\bibitem{fredrickson2006collective}
Fredrickson E~D, Bell R~E, Darrow D~S, Fu G~Y, Gorelenkov N~N, LeBlanc B~P,
  Medley S~S, Menard J~E, Park H, Roquemore A~L, Heidbrink W~W, Sabbagh S~A,
  Stutman D, Tritz K, Crocker N~A, Kubota S, Peebles W, Lee K~C and Levinton
  F~M 2006 {\em Physics of Plasmas\/} {\bf 13} 056109
  \urlprefix\url{https://doi.org/10.1063/1.2178788}

\bibitem{gryaznevich2006perturbative}
Gryaznevich M~P and Sharapov S~E 2006 {\em Nuclear Fusion\/} {\bf 46}
  S942--S950 \urlprefix\url{https://dx.doi.org/10.1088/0029-5515/46/10/S11}

\bibitem{gryaznevich2008recent}
Gryaznevich M~P, Sharapov S~E, Lilley M, Pinches S~D, Field A~R, Howell D,
  Keeling D, Martin R, Meyer H, Smith H, Vann R, Denner P, Verwichte E and {the
  MAST team} 2008 {\em Nuclear Fusion\/} {\bf 48} 084003
  \urlprefix\url{https://dx.doi.org/10.1088/0029-5515/48/8/084003}

\bibitem{bierwage2017self}
Bierwage A, Shinohara K, Todo Y, Aiba N, Ishikawa M, Matsunaga G, Takechi M and
  Yagi M 2017 {\em Nuclear Fusion\/} {\bf 57} 016036
  \urlprefix\url{https://dx.doi.org/10.1088/1741-4326/57/1/016036}

\bibitem{melnikov2018detection}
Melnikov A~V, Ascasibar E, Cappa A, Castej\'{o}n F, Eliseev L~G, Hidalgo C,
  Khabanov P~O, Kharchev N~K, Kozachek A~S, Krupnik L~I, Liniers M, Lysenko
  S~E, de~Pablos J~L, Sharapov S~E, Ufimtsev M~V, Zenin V~N and {TJ-II Team}
  2018 {\em Nuclear Fusion\/} {\bf 58} 082019
  \urlprefix\url{https://dx.doi.org/10.1088/1741-4326/aabcf8}

\bibitem{fulop2009magnetic}
F\"{u}l\"{o}p T, Smith H~M and Pokol G 2009 {\em Physics of Plasmas\/} {\bf 16}
  022502 \urlprefix\url{https://doi.org/10.1063/1.3072980}

\bibitem{shi2022solenoid}
Shi Y, Liu B, Song S, Song Y, Song X, Tong B, Cheng S, Liu W, Wang M, Sun T,
  Guo D, Li S, Li Y, Chen B, Gu X, Cai J, Luo D, Banerjee D, Zhao X, Yang Y,
  Luo W, Zhou P, Wang Y, Ishida A, Maekawa T, Liu M, Yuan B, Peng Y~K~M and
  {the EXL-50 Team} 2022 {\em Nuclear Fusion\/} {\bf 62} 086047
  \urlprefix\url{https://dx.doi.org/10.1088/1741-4326/ac71b6}

\bibitem{li2021quasi}
Li S~J, Bai R~H, Tao R~Y, Li N, Lun X~C, Liu L~C, Liu Y, Liu M~S and Deng B~H
  2021 {\em Journal of Instrumentation\/} {\bf 16} T08011
  \urlprefix\url{https://dx.doi.org/10.1088/1748-0221/16/08/T08011}

\bibitem{wang2023radio}
Wang M, Lun X, Bo X, Liu B, Liu A and Shi Y 2023 {\em Plasma Science and
  Technology\/} {\bf 25} 045104
  \urlprefix\url{http://pst.hfcas.ac.cn/EN/10.1088/2058-6272/aca373}

\end{thebibliography}

\end{document}